\documentclass[preprints,article,accept,moreauthors,pdftex]{Definitions/mdpi}
\usepackage{bm}
\firstpage{1}
\pubvolume{1}
\issuenum{1}
\articlenumber{0}
\pubyear{2021}
\copyrightyear{2020}
\datereceived{}
\dateaccepted{}
\datepublished{}
\hreflink{https://doi.org/} % If needed use \linebreak
\Title{Scalable codes for precision calculations of properties of complex atomic systems}

\TitleCitation{Scalable codes for precision calculations of properties of complex atomic systems}

\Author{Charles Cheung $^{1*}$\orcidA{}, Marianna Safronova $^{1}$\orcidB{}, and Sergey Porsev $^{1,2}$\orcidC{}}

\AuthorNames{Charles Cheung, Marianna Safronova and Sergey Porsev}

\AuthorCitation{Cheung, C.; Safronova, M.; Porsev, S.}
\address{%
$^{1}$ \quad Department of Physics and Astronomy, University of Delaware, DE, USA\\
$^{2}$ \quad Petersburg Nuclear Physics Institute of NRC "Kurchatov Institute", Gatchina, Leningrad District, 188300, Russia}

\corres{Correspondence: ccheung@udel.edu; }

\abstract{High precision atomic data is indispensable for studies of fundamental symmetries and tests of fundamental physics postulates, development of atomic clocks and other ultracold atom experiments, astrophysics, plasma science, and many others. We developed new parallel atomic structure code package that enabled computations that were not possible before due to system complexity and allowed much quicker computations with higher accuracy for simpler systems. We achieved  near-perfect linear scalability and efficiency with the number of cores paving the way towards the future where we finally be able to treat most open-shell systems with good accuracy. We present several examples of new capabilities, correlating all 60 electrons in the highly charged Ir$^{17+}$ ion and calculations predicting the $3C/3D$ line intensity ratio in Fe$^{16+}$. }

\begin{document}
%%%%%%%%%%%%%%%%%%%%%%%%%%%%%%%%%%%%%%%%%%
%\setcounter{section}{-1} %% Remove this when starting to work on the template.
%\section{How to Use this Template}

%The template details the sections that can be used in a manuscript. Note that the order and names of article sections may differ from the requirements of the journal (e.g., the positioning of the Materials and Methods section). Please check the instructions on the authors' page of the journal to verify the correct order and names. For any questions, please contact the editorial office of the journal or support@mdpi.com. For LaTeX-related questions please contact latex@mdpi.com.
%The order of the section titles is: Introduction, Materials and Methods, Results, Discussion, Conclusions for these journals: aerospace,algorithms,antibodies,antioxidants,atmosphere,axioms,biomedicines,carbon,crystals,designs,diagnostics,environments,fermentation,fluids,forests,fractalfract,informatics,information,inventions,jfmk,jrfm,lubricants,neonatalscreening,neuroglia,particles,pharmaceutics,polymers,processes,technologies,viruses,vision

\section{Introduction}
Studies of the fundamental symmetries with atoms and ions require knowledge of atomic properties as well as calculations to extract potential new physics from the experiments \cite{SafBudDem18}. For example, the most accurate atomic calculation for a heavy atom was carried out to study Cs parity violation \cite{PorBelDer09,PorBelDer10}. A precision (0.35\% accuracy) experiment  \cite{WooBenCho97} had to be supplemented by theory computation of a comparable accuracy in order to extract the value of the weak charge and test the standard model of elementary particles. The theory accuracy is more than a order of magnitude worse in Yb, not sufficient for a similar analysis of the Yb experiments \cite{2009PhRvL.103g1601T,Antypas_2018} motivating studies with multiple isotopes \cite{Antypas_2018}.  While Yb has two valence electrons, low-lying excitations from the closed shells that cannot so far be accurately treated by any high-precision methods significantly lower attainable precision.  Theory accuracy is even worse for more compacted systems such as Dy \cite{leefer2014new} or Sm of interest to parity violation and other studies. Studies of CP-violation with atoms also require computations of the enhancement factors \cite{2012PhRvL.108q3001P}. Studies of Lorentz symmetry violations need computations of the Lorentz invariance violating matrix elements \cite{2019Natur.567..204S}. Moreover, development of experiments with new systems requires accurate knowledge of many atomic properties. For example, development of even more precise atomic clocks \cite{2015RvMP...87..637L} that are used to test the Einstein's equivalence principle, search for the variation of the fundamental constants and dark matter  \cite{SafBudDem18} is accelerated by strong theory-experimental collaborations. 

In the present work we improve on the capabilities of an \textit{ab initio} atomic structure code package for calculating atomic properties of complex many-electron systems. The methods used here, including configuration interaction (CI) and the combination of CI with either many body perturbation theory (CI+MBPT) or the linearized coupled cluster method (CI+all-order), are very broadly applicable to any atom in the periodic table. Numerous problems that were not tractable before have been solved with our newly developed parallel codes, including problems in astrophysics, metrology of highly charged ions, neutral atoms, and negative ions. Our focus will be on the computational developments that enable these new large-scale calculations.

A version of the CI+MBPT code package was modified for public use and published in Computer Physics Communications in 2015 by M. Kozlov \textit{et al} \cite{KozPorSaf15}. Although not published yet, the inclusion of the all-order method which provides accurate solutions for a large number of properties of atoms and ions with up to 5 -- 6 valence electrons has been completed and made fully compatible with this code package. The main computational developments involved implementing the message passing interface (MPI) to parallelize computationally expensive portions of the programs. This allows us to fully take advantage of modern high performance computing facilities, such as the University of Delaware high performance Caviness community cluster, where we developed and tested our programs. We use the new package of parallel programs to consider two cases of particular experimental significance: Ir$^{17+}$, which was proposed for the development of novel atomic clocks, and Fe$^{16+}$, which has lines essential for plasma diagnostics tools for astrophysics. We found great success in the parallelization efforts, as new programs enabled precision calculations of these systems beyond what was desired. 

%%%%%%%%%%%%%%%%%%%%%%%%%%%%%%%%%%%%%%%%%%

%Materials and Methods should be described with sufficient details to allow others to replicate and build on published results. Please note that publication of your manuscript implicates that you must make all materials, data, computer code, and protocols associated with the publication available to readers. Please disclose at the submission stage any restrictions on the availability of materials or information. New methods and protocols should be described in detail while well-established methods can be briefly described and appropriately cited.
%
%Research manuscripts reporting large datasets that are deposited in a publicly available database should specify where the data have been deposited and provide the relevant accession numbers. If the accession numbers have not yet been obtained at the time of submission, please state that they will be provided during review. They must be provided prior to publication.

\section{Theory and Methods}
In this section, we describe the methods used in our programs and applications. For any many-electron system, we can divide all electrons into core and valence electrons. In this way, we can separate the electron-electron correlation problem into one describing the valence-valence correlations under the frozen-core approximation, and another describing the core-core and core-valence correlations. In the initial approximation, we start from the solution of the restricted Dirac-Hartree-Fock (HFD) equations in the central field approximation to construct one-electron orbitals for the core and valence electrons. Virtual orbitals can be constructed from B-splines or by other means to account for correlations. The valence-valence correlation problem is solved using the CI method, while core-core and core-valence correlations are included using either MBPT or the all-order method. In either case, we form an effective Hamiltonian in the valence CI space, then diagonalize the effective Hamiltonian using the CI method to find energies and wave functions for the low-lying states.

\subsection{The CI Method}\label{sec:ci}
The CI method is a standard \textit{ab initio} method for calculating atomic properties of many-electron systems. In the valence space, the CI wave function is constructed as a linear combination of all distinct states of a specified angular momentum $J$ and parity
\begin{equation}
	\psi=\sum_ic_i\Phi_i,
\end{equation}
where the set $\left\{\Phi_i\right\}$ are Slater determinants generated by exciting electrons from some reference configurations 
%obtained from HFD theory 
to higher orbitals. 

Varying the coefficients $c_i$ results in a generalized eigenvalue problem
\begin{equation}
	\sum_j\langle\Phi_i|H|\Phi_j\rangle c_j = Ec_i,
\end{equation}
which can be written in matrix form and diagonalized to find the energies and wave functions of the low-lying states. The energy matrix of the CI method can be
obtained as a projection of the exact Hamiltonian $H$ onto the CI subspace $H^\text{CI}$~\cite{Dzuba1996}
%The CI wave function is then used to solve the time-independent Schr\"odinger equation 
%\begin{equation}
%	H\Phi_n = E_n\Phi_n.
%\end{equation} 
%The CI Hamiltonian can be written as
\begin{equation}\label{eq:hci}
	H^\text{CI}=E_\text{core}+\sum_{i>N_\text{core}}h_i^\text{CI}+\sum_{j>i>N_\text{core}}V_{ij},
\end{equation}
where $E_\text{core}$ is the energy of the frozen core, $N_\text{core}$ is the number of core electrons, $h_i^\text{CI}$ accounts for the kinetic energy of the valence electrons and their interaction with the central field, and $V_{ij}$ accounts for the valence-valence correlations.

Having obtained from CI the many-electron states $|J M\rangle$ and $|J' M'\rangle$ with the total angular momenta $J,J'$ and their projections $M,M'$, one can form density transition matrix in terms of the one-electron states  $|nljm\rangle$ \cite{KozPorSaf15}
\begin{equation}
	\hat{\rho}=\rho_{nljm,n^\prime l^\prime j^\prime m^\prime}|nljm\rangle\langle n^\prime l^\prime j^\prime m^\prime|,
\end{equation}
where
\begin{equation}
	\rho_{nljm,n^\prime l^\prime j^\prime m^\prime}=\langle J^\prime M^\prime|a_{n^\prime l^\prime j^\prime m^\prime}^\dagger a_{nljm}|JM\rangle. 
\end{equation}
Here un-primed indices refer to the initial state and primed indices refer to the final state. The many-electron matrix element can then be written as 
\begin{equation}
	\langle J^\prime M^\prime|T_q^L|JM\rangle=\text{Tr}\,\rho_{nljm,n^\prime l^\prime j^\prime m^\prime}\langle n^\prime l^\prime j^\prime m^\prime|T_q^L|nljm\rangle,
\end{equation}
where the trace sums over all quantum numbers $(nljm)$ and $(n^\prime l^\prime j^\prime m^\prime)$, and $T_q^L$ is the spherical component of the tensor operator of rank $L$. Using the Wigner-Eckart theorem, one can reduce the many-electron matrix element to
\begin{equation}\label{eq:rdtme}
	\langle J^\prime \Vert T^L \Vert J\rangle = \text{Tr}\,\rho_{nlj,n^\prime l^\prime j^\prime}^L \langle n^\prime l^\prime j^\prime\Vert T^L \Vert nlj\rangle,
\end{equation}
where
\begin{equation}\label{eq:rdtm}
\rho_{nlj,n^\prime l^\prime j^\prime}^L = (-1)^{J^\prime -M^\prime}\left(
\begin{array}{ccc}
	J^\prime & L & J \\ -M^\prime & q & M
\end{array}\right)^{-1} \sum_{mm^\prime} (-1)^{j^\prime-m^\prime}\left(
\begin{array}{ccc}
	j^\prime & L & j \\ -m^\prime & q & m
\end{array}\right) \rho_{nljm,n^\prime l^\prime j^\prime m^\prime}.
\end{equation}

We have developed new parallel programs based on these methods: \texttt{conf} realizes the CI method, which forms the CI Hamiltonian and uses Davidson's algorithm of diagonalization \cite{Davidson1975} to find low-lying energies and wave functions; \texttt{dtm} calculates reduced matrix elements (\ref{eq:rdtme}) of one-electron operators by forming the reduced density transition matrices (\ref{eq:rdtm}). We discuss the challenges of developing the parallel programs in Sec.~\ref{sec:compdev}.

\subsection{Selecting important configurations with valence perturbation theory (CI+PT)}\label{sec:ci+pt}

As the number of configurations contributing to the CI wave function grows exponentially with the number of valence electrons, efficient selection of the most important configurations from a set of configurations becomes the main challenge of accurate computations. To significantly reduce the number of configurations, we further developed a method suggested in Ref.~\cite{RKP01e} to predict important configurations based on a set of configurations with known weights. This method can be used to optimize the CI space by identifying the most important configurations from a list of CI configurations using perturbation theory (PT) \cite{RKP01e}. 

All second-order corrections are taken into account and added to the energy calculated from CI to obtain the total energy, $E^\text{CI}=E_0+E_1$, while first-order corrections to the wave functions are stored for use in subsequent CI calculations. This process of using CI on a small subspace, calculating corrections via PT, and reordering the list of configurations in descending weights can be repeated several times to form the most optimal CI subspace. Once the energy differences between subsequent CI calculations are relatively small, it can be assumed that convergence has been met. 

We've developed a new parallel program \texttt{conf\_pt} that realizes the CI+PT method. The parallel version enables computations of extremely large problems, with tests running up to 400 million determinants. 

\subsection{Including core correlations with other methods (CI+MBPT/CI+all-order method)}\label{sec:core}
The CI+MBPT~\cite{Dzuba1996,DzuKozPor98,SafKozJoh09} and CI+all-order~\cite{SafKozJoh09} methods include core-core and core-valence correlations using the combination of CI with second-order MBPT and the linearized coupled cluster method, respectively. The many-electron Schr\"odinger equation can be written as
\begin{equation}\label{eq:effse}
	H_\text{eff}\Psi=E\Psi,
\end{equation}
where the effective Hamiltonian has the form
\begin{equation}
	H_\text{eff}=H^\text{CI}+\Sigma.
\end{equation}
Here $H^\text{CI}$ is the CI Hamiltonian described by Eq.~\ref{eq:hci} and $\Sigma=\Sigma_1+\Sigma_2$ is the core-valence correlation potential obtained from either MBPT or the all-order method, where $\Sigma_1$ and $\Sigma_2$ are the one- and two-electron parts of the core-valence correlation potential, respectively. Eq.~\ref{eq:effse} is solved by diagonalizing the effective Hamiltonian using the CI method to obtain energies and wave functions of the low-lying states, then matrix elements can be calculated as described in Sec.~\ref{sec:ci}.

%%%%%%%%%%%%%%%%%%%%%%%%%%%%%%%%%%%%%%%%%%
\section{Computational developments}\label{sec:compdev}
\begin{figure*}[thb!]
	\centering
	\includegraphics[width=\textwidth]{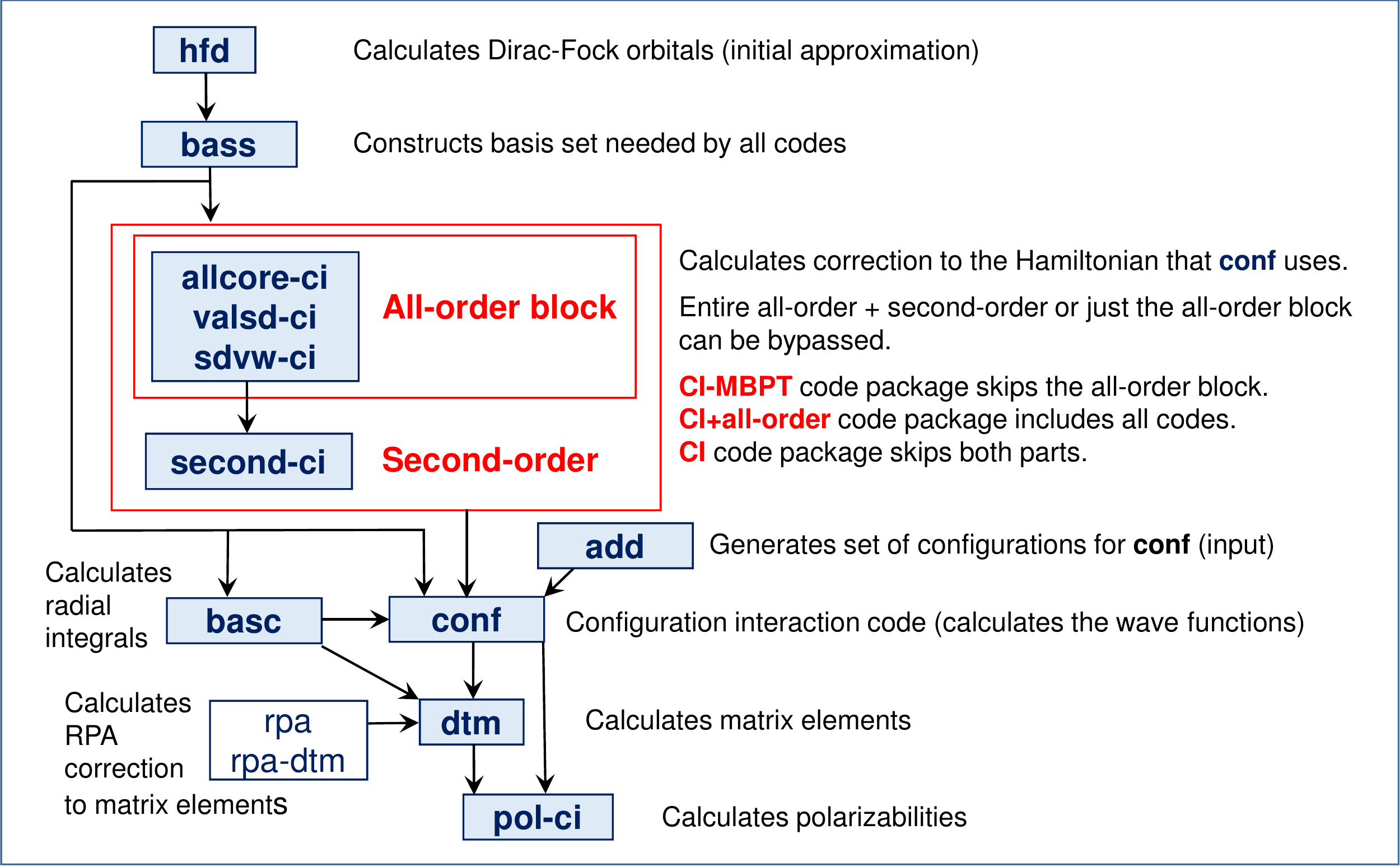}
	\caption{The scheme of the CI/CI+MBPT/CI+all-order code package.}
	\label{fig:codes}
\end{figure*}

The complete CI/CI+MBPT/CI+all-order code package scheme is illustrated in Fig.~\ref{fig:codes}. Of the programs listed in the scheme, the CI program (\texttt{conf}) and the matrix elements program (\texttt{dtm}) have been parallelized with the message passing interface (MPI) \cite{Gropp2014,Gropp2015}. The CI+PT program (\texttt{conf\_pt}) is an auxiliary program that has also been parallelized with MPI. There are several key features of our recently developed parallel codes over previous serial (non-parallel) programs published in Ref.~\cite{KozPorSaf15}. For all parallel programs, we focused on improving readability and usability, reducing the total memory footprint, implementing dynamic memory allocation where possible, and parallelizing the computationally expensive portions of code using MPI. 

The newly developed parallel programs enable computations that were not possible before due to lack of memory or prohibitive computational times. For smaller systems, we can now do calculations with much more accuracy and much faster, allowing a broader range of effects to be investigated in a short period of time. With the original serial version of the programs, calculating properties of very small systems of 1 -- 2 valence electrons could take up to a few days, while runs for more complex systems with 3 -- 6 valence electrons could last up to weeks. Some problems involving systems with more than 6 valence electrons were simply intractable with this code due to lack of memory or prohibitively long computation times. One of the main objectives of parallelizing the programs was to reduce the computational time required for calculations of properties of complex atomic systems. 

\subsection{Parallelization of codes}
Our main focus will be on the developments of the parallel CI, CI+PT, and density transition matrix programs. Each of these methods require calculating matrix elements between a pair of determinants, which is very difficult to parallelize due to the nature of the problem. In the CI program, the CI Hamiltonian matrix as well as the matrix of the operator $\bm{J^2}$ are constructed, in the CI+PT program, the CI and PT blocks of the Hamiltonian matrix are constructed, and in the density transition matrix elements program, the density or transition matrix is constructed. Construction of each of these matrices are essentially a different variant of the same problem. 

We will describe only the parallel implementation to the CI program, but the methods discussed here are applicable in general to the other parallel programs. The main difficulty here is the intrinsic unpredictability of the computational method, i.e. one cannot guess how many non-zero matrix elements there are without explicitly comparing all pairs of determinants. When forming the matrices and calculating the matrix elements, one has to compare electron occupancies between determinants for each configuration. The Slater-Condon rules \cite{Slater1929,Condon1930} describe the number of operations that are done for each matrix element depending on the number of differences between the pair of determinants forming the matrix element. 

The construction of the matrix follows a two-loop structure: an outer loop iterates over the total number of determinants and an inner loop iterates over the total number of configurations. Determinants belonging to each configuration are paired and the value of the matrix element is calculated if there are less than 3 differences between the pair of determinants. We separate the computation into two stages: (i) a comparison stage to find pairs of determinants with less than 3 differences between them and (ii) a calculation stage to calculate the value of the matrix element. After all determinants have been compared, one final loop is done through the total number of non-zero matrix elements to calculate their values. With this implementation, we have achieved near-perfect linear scalability and efficiency in our tests with up to 550 computing cores. It is also possible, and much simpler, to distribute the workload by dividing the outer loop by the number of cores. However, this results in very uneven load-balancing which is exacerbated for larger computations. This is not the case when dealing with smaller matrices such as in the case of the density and transition matrices. 

In the serial implementation, non-zero matrix elements were written to disk due to lack of memory of the computers used when the codes were first developed. However with modern computers, we are able to store the non-zero matrix elements in memory, which also has the advantage of reducing computation time due to the removal of file input and output (I/O) to disk. The serial program wrote each matrix element to disk with 24 bytes: 8 bytes for the counter, 4 bytes for each index, and 8 bytes for the value of the non-zero matrix element. The parallel program removes the redundant 8 bytes for the id, reducing the memory requirement of storing the Hamiltonian matrix by 33\%.

Since the matrix elements of the CI Hamiltonian are distributed across cores, the Davidson procedure had to be modified to take advantage of the parallelization. The Davidson procedure diagonalizes the CI Hamiltonian matrix to find energies and wave functions of the low-lying states. In the present code, only the calculation of matrix-vector products have been parallelized. Since each core holds a portion of the Hamiltonian matrix in memory, the Davidson algorithm had to be modified so each core is only responsible for computing products of their stored matrix elements with the wave function. 

\subsection{Speedup of parallel programs}
The performance of the parallel \texttt{conf} program was tested using the Ir$^{17+}$ ion. We describe Ir$^{17+}$ as the ion with 30 valence electrons in the open $4f$, $4d$, and $4p$ shells with an [$8spdfg$] basis set. The basis set is designated by the highest principal quantum number for each partial wave included. For example, the [$8spdfg$] basis set includes $1-8s$, $2-8p$, $3-8d$, $4-8f$, and $5-8g$ orbitals. These test calculations included 24\,895 relativistic even-parity configurations and 17\,431\,323 determinants. As reference, the total computation time of the original serial calculation just short of 2 weeks. 

\begin{specialtable}
	\caption{\label{tab:conf_spdups}
		The runtime of subroutines \texttt{FormH} and \texttt{Diag4} of the parallel \texttt{conf} program for increasing number of compute cores and the speedups of the parallel code are presented in seconds (s) for increasing number of cores N, relative to the code ran with 50 cores (N=50). These large test runs were done with Ir$^{17+}$ with 24\,895 relativistic configurations and $17.4\times10^6$ determinants. Note that the total times include all serial subroutines outside of \texttt{FormH} and \texttt{Diag}. }
	\centering
	\begin{tabular}{ l  c c c  c c c}
		\toprule
		\multicolumn{1}{c}{} &
		\multicolumn{3}{c}{runtime (s)} &
		\multicolumn{3}{c}{speedup (from N=50)}\\
		\midrule
		N   & \texttt{FormH}  & \texttt{Diag4} & total & \texttt{FormH} & \texttt{Diag4} & total \\ 
		\midrule
		50  & 22571  & 1343      & 24218  & 1  &  1 &  1 \\  
		100 & 12843  & 1514      & 14593  & 1.8  & 0.9 &  1.7  \\ 
		200 & 5800   & 957     & 6927   & 3.9 & 1.4 & 3.5 \\ 
		300 & 3810 & 678     & 4610 & 5.9 & 2.0 & 5.3  \\ 
		400 & 2913 & 596     & 3646 & 7.8 & 2.3 & 6.6 \\ 
		500 & 2292 & 535     & 2958 & 9.9 & 2.5 & 8.2 \\ 
		\bottomrule
	\end{tabular}
\end{specialtable}

In Table~\ref{tab:conf_spdups}, we compare runtimes of the formation of the CI Hamiltonian matrix (\texttt{FormH}) and the Davidson iterative procedure (\texttt{Diag4}), as well as the total computational time, for increasing number of computing cores. Comparing large number of cores to the base \texttt{N=50} case, we see that there is a near-perfect linear scalability up to 500 cores for the \texttt{FormH} subroutine, but see very minimal performance gain for \texttt{Diag} from 200 to 500 cores. 

The Davidson procedure of our parallel CI program does not scale well since only the calculation of matrix-vector products was parallelized, and a large majority of the procedure remains serial. Since the Davidson procedure typically does not run as long as the formation of the Hamiltonian, the performance of the Davidson procedure was deemed sufficient for our problems, leaving remaining optimizations to a future project. 

The efficiency of the total speedup attained with the parallel CI program is about 70 -- 80\% with the number of cores. The speedup achieved with the parallel CI+PT program and the parallel matrix element program is very similar, with the parallel CI+PT program averaging around 75 -- 85\% efficiency and the parallel matrix element program averaging around 80 -- 90\% efficiency with the number of cores. Since the development of these parallel programs, they have been routinely used in our group to calculate various atomic properties of many-electron systems.

\subsection{Selection of important configurations}
In addition to the parallel codes, we've developed algorithms and auxiliary codes for significantly reducing the number of configurations by efficiently identifying dominant configurations from a list of configurations. It is necessary to be able to select the most important configurations for the valence CI space when the dimensionality of the CI problem becomes intractably large. 

The first algorithm involves using the CI+PT method to predict important configurations based on a set of already selected configurations with known weights, as discussed in Sec.~\ref{sec:ci+pt}. Initially, a configuration list with no weights is constructed by allowing excitations from a few basic configurations. A small subspace is chosen from the top of the list to generate initial weights for each of the chosen configurations in the list using CI, then weights of all other configurations are calculated using PT. We then reorder the initial configuration list by descending weights, then repeat this process until convergence is met, i.e. the CI space has been saturated as it has taken into account the most important configurations. The convergence is checked by comparing energy levels computed from consecutive CI and PT procedures. 

The second algorithm explores the importance of the configurations based on the correlations between electrons that are present in the configurations. We find that contributions of electrons in orbitals of the same partial wave are very regular. For example, if our calculations of Fe$^{16+}$ show that the contribution of the $1s^22s2p^56s^2$ configuration is negligible, then all similar configurations where the two last electrons have higher principle quantum number, e.g. $1s^22s2p^56s7s$, $1s^22s2p^56s8s$, $1s^22s2p^57s^2$, can be omitted. However, if the $1s^22s2p^66f$ configuration has a large weight, then other configurations with $7f$ and $8f$ electrons should be included.

%%%%%%%%%%%%%%%%%%%%%%%%%%%%%%%%%%%%%%%%%%

\section{Applications}

\subsection{Optical clocks based on highly charged ions}
Highly charged ions (HCI) such as Ir$^{17+}$ are of particular interest for the development of novel atomic clocks due to its very high sensitivity to the variation of the fine structure constant $\alpha$ and related dark matter searches \cite{Crespo2008,BDF10,BDFO11,KozSafCre18}. There are many advantages to creating optical clocks with HCI, such as enhanced sensitivity to the variation of $\alpha$, heavily suppressed systematic effects, and estimated potential clock uncertainty beyond the $10^{-18}$ limit \cite{BerDzuFla12,DerDzuFla12,DzuDerFla12a,DzuDerFla13E}. Recent developments in quantum logic techniques for HCI spectroscopy have made rapid progress in the development of HCI possible \cite{Schmoger2015}.

In the case of Ir$^{17+}$, theoretical calculations are particularly difficult due to atomic configurations with holes in the $4f$ shell leading to a very large uncertainty in prior theoretical results, with over 50\% for the predicted clock transition energy \cite{WCB15,SFS15}. No clock transitions or any $E1$ transitions have been found despite over 6 years of experimental effort. These transitions were expected to be observed in recent experiments since their predicted transition rates were well within experimental capabilities; especially since the $M1$ transitions with much smaller transition rates have been observed. The lack of observations for the $E1$ transitions brought serious concerns over the accuracy of theoretical predictions, even to the point of doubt of approximate spectral range. With our newly developed parallel programs, we resolved this problem and for the first time definitively demonstrated the ability to converge CI in systems with a few holes in the $4f$ shell and place uncertainty bounds on our results. Our results explain the lack of observations of the $E1$ transitions and provide a pathway towards detection of clock transitions based on highly charged ions. Here we will only summarize the results of our application of the new parallel code to Ir$^{17+}$. A detailed discussion can be found in Ref.~\cite{CheSafPor20}. 

\begin{specialtable*}[thb!]
		\caption{\label{tab:ir17_energies} Energies of Ir$^{17+}$ (in cm$^{-1}$) obtained using CI with increasing number of open shells. Only configurations obtained by exciting $4f$ and $5s$ electrons are included in the ``$5s4f$ only'' column. Contributions from exciting electrons from the $4d$ shell are given in the column labeled ``$4d$ contr.'', and contributions of all other shells are given separately in the next columns. The results with all 60 electrons correlated by the CI are listed in the column ``All shells open''. Sum of all other corrections is given in the column labeled ``Other'' - see Ref. \cite{CheSafPor20} for detailed explanation. }
		\begin{tabular}{lcccccccccccc} 
			\toprule
			\multicolumn{2}{c}{Configuration}&
			\multicolumn{1}{c}{$5s4f$ only}&
			\multicolumn{1}{c}{$4d$}& 
			\multicolumn{1}{c}{$4p$}&
			\multicolumn{1}{c}{$4s$}& 
			\multicolumn{1}{c}{$3d$}&  
			\multicolumn{1}{c}{$1s2s3s$}&
			\multicolumn{1}{c}{$3p$}&
			\multicolumn{1}{c}{$2p$}& 
			\multicolumn{1}{c}{All shells} &
			\multicolumn{1}{c}{Other}&  
			\multicolumn{1}{c}{Final}\\
			\multicolumn{3}{c}{}&
			\multicolumn{1}{c}{contr.} &\multicolumn{1}{c}{contr.} &\multicolumn{1}{c}{contr.} &\multicolumn{1}{c}{contr.} &\multicolumn{1}{c}{contr.} &\multicolumn{1}{c}{contr.} &\multicolumn{1}{c}{contr.} &\multicolumn{1}{c}{open}& \multicolumn{2}{c}{}\\
			\midrule
			$4f^{13}5s$   & $^3\!F^o_4$ &  0       &      0 &     0   &      0  &     0     &   0    &    0   &0   &   0    & 0   &0 \\
						  & $^3\!F^o_3$ & 4714     &     31 &    15   &    14   &     8     &  -3    &    2   &0   & 4781   &-4   &4777 \\
						  & $^3\!F^o_2$ & 25170    &    -75 &    14   &    13   &    75     &  -2    &   25   &-4  & 25220  &-34  &25186 \\ [0.5pc]
%						  & $^1\!F^o_3$ & 30137    &    116 &    51   &    33   &    73     &  -3    &   23   &-4  & 30426  &-31  &30395 \\ [0.5pc]
			$4f^{14}$     & $^1\!S_0$   & 9073     &   5797 &  -931   &  -1994  &   1097    & -240   &   -54  &9   &  12757 &-375 &12382 \\ [0.5pc]
			$4f^{12} 5s^2$& $^3\!H_6$   & 36362    &  -8549 &   460   &   1848  &   -403    &  183   &   294  &144 &  30339 &-56  &30283 \\
						  & $^3\!F_4$   & 46303    &  -8680 &    -5   &   1858  &   -410    &  184   &   251  &144 &  39645 &-81  &39564 \\
						  & $^3\!H_5$   & 59883    &  -8638 &   454   &   1858  &   -326    &  183   &   324  &143 &  53882 &-84  &53798 \\
						  & $^3\!F_2$   & 68786    &  -8751 &  -188   &   1690  &   -384    &  253   &   191  &64  &  61662 &-233 &61429  \\
%						  & $^1\!G_4$   & 69099    &  -9043 &   165   &   1868  &   -322    &  184   &   304  &143 &  62397 &-136 &62261  \\
%						  & $^3\!F_3$   & 71963    &  -8894 &   146   &   1836  &   -332    &  179   &   266  &146 &  65309 &-129 &65180  \\
%						  & $^3\!H_4$   & 91038    &  -8784 &    78   &   1894  &   -245    &  187   &   340  &142 &  84650 &-126 &84524  \\
%						  & $^1\!D_2$   & 97473    &  -9618 &  -110   &   1735  &   -334    &  270   &   177  &48  &  89639 &-366 &89273  \\
%						  & $^1\!J_6$   & 109332   &  -10201&   268   &   1809  &   -304    &  171   &   212  &150 &  101437&-301 &101136  \\
			%             & $^3\!P_2$   & 123129   &  -9788 &   -348  &   1646  &   -273    &  308   &   168  & 2  &  114846&     & 114846 \\
			\bottomrule
	\end{tabular}
\end{specialtable*}

We find that the best initial approximation is achieved by solving restricted DHF equations with partially filled shells, namely $[1s^2\dots 4d^{10}]4f^{13}5s$. The hybrid approaches described in Sec.~\ref{sec:core} that incorporates core excitations into the CI by constructing an effective Hamiltonian using MBPT or the all-order method can not be used with such an initial approximation. Therefore, we treat the inner shells with the CI method, leading to an exponential increase in the number of configurations. We found that while the weights of most configurations are small, the number of important configurations were very large. 

Our new parallel programs allowed us to increase the valence space from 14 electrons to all 60, and to include 250\,000 configurations, resulting in 133 million Slater determinants, a factor of 20 increase to what was previously feasible. In order to definitively ensure the reliability of the theoretical calculations, we considered all possible contributions that may affect the accuracy of the computations and ensure the convergence in all numerical parameters, including the number and type of configurations included in the CI, the size of the orbital basis set used to construct CI configurations, inclusion of the quantum electrodynamics (QED) corrections, and inclusion of the Breit corrections beyond the Gaunt term. We found that by far the largest effect comes from the inclusion of the inner electron shells into the CI. 

We start with the most straightforward CI computation that includes single and double excitations from the $4f$ and $5s$ valence shells, similar to Ref~\cite{BDFO11}. The excitations are allowed to each of the basis set orbitals up to $[7spdfg]$. Next, we allow all $4d$ electrons into the valence space and allow excitations of any of the 24 electrons from the $4d^{10}4f^{13}5s$ shells to the same basis set orbitals up to $[7spdfg]$. We find drastic changes in the energies of the $E1$ transitions when excitations are allowed from the $4d$ shell. Due to such large contributions, we continued to include more and more electrons of the inner shells into the CI valence space, until all 60 electrons have been included. With the same $[7spdfg]$ basis set, both single and double excitations are allowed from the $4f, 4d, 4p, 4s$ and $3d$ shells, and only single excitations are included for all other shells. We tested that the double excitation contribution is small for these inner shells and can be omitted at the present level of accuracy. The results, obtained with different number of shells included in the CI valence space, are given in Table~\ref{tab:ir17_energies}. Contributions from increased size of the basis set, triple excitations, full Breit, and QED were found to be small at the present level of accuracy. Three basis sets of increasing sizes $[7spdfg]$, $[8spdfg]$, and $[10spdfg]$ were used to test basis set convergence. 

While allowing excitations from the $4d$ shell drastically reduced the energies of the $E1$ transitions, allowing excitations from the $4p$ shell drastically reduced several $E1$ matrix elements and rates to well below the detector ability. We have identified several other transitions at different wavelengths for the future $E1$ transition search. As soon as any $E1$ transition is detected, we will be able to obtain a better prediction for the proposed clock transition wavelength. 

The largest Ir$^{17+}$ run with the latest version of the parallel CI program included 96\,622 configurations and 58\,224\,918 determinants, while the largest number of determinants included in the old serial CI program was about 5\,000\,000. The largest run calculated and stored over 100 billion Hamiltonian matrix elements, with the whole run requiring a total of 2\,880 GB of memory, which was the maximum amount available to us at the time. Distributed across 80 cores, this run took 2 days and 18 hours to complete. A larger number of cores were not used due to the large amount of memory required to store the basis set, the Hamiltonian matrix, and subsequent arrays for the Davidson procedure. 

\subsection{Calculation of the 3C/3D line intensity ratio in Fe XVII}
Some of the brightest lines of the spectra of many hot astrophysical objects arise from Fe$^{16+}$ around 15 \AA, namely the resonance line $3C$ $([(2p^5)_{1/2}3d_{3/2}]_{J=1}\rightarrow[2p^6]_{J=0})$ and the intercombination line $3D$ $([(2p^5)_{3/2}3d_{5/2}]_{J=1}\rightarrow[2p^6]_{J=0})$. They are crucial for plasma diagnostics of electron temperatures, elemental abundances, ionization conditions, velocity turbulences, and opacities. For the last four decades, there has been a persistent discrepancy between the observed intensity ratios and advanced plasma models, diminishing the utility of high-resolution X-ray observations. A recent experiment measured the most accurate $3C/3D$ oscillator strength to date, in an attempt to explain this puzzle \cite{Kuehn2020}. 

We carried out a precision calculation with our newly developed MPI version of the CI program, allowing us to drastically increase the number of included configurations to over 230\,000. This calculation correlated all 10 electrons, included full Breit and QED corrections, and predicted the transition rates with 1 -- 2\% accuracy. Our calculations ruled out incomplete inclusion of the electronic correlations in theoretical calculations as the potential explanation of the puzzle. A detailed study of the latest experiment and theoretical work done can be found in Ref.~\cite{Kuehn2020}. 

\begin{specialtable*}[htb!]
	\caption{\label{fetable} Contributions to the energies (in cm$^{-1}$) of Fe$^{16+}$ calculated with increased basis set sizes and number of configurations. Contributions from triple excitations, excitations from the $1s^2$ shell, and QED contributions are given in their respective columns. The last line gives the $3C-3D$ energy difference in eV.}
	\centering
		\begin{tabular}{lccccccccc}
			\toprule
			\multicolumn{2}{c}{Configuration}&
			\multicolumn{1}{c}{$[5spdf6g]$}&
			\multicolumn{1}{c}{Triples}& 
			\multicolumn{1}{c}{$1s^2$}&
			\multicolumn{1}{c}{$+[12spdfg]$}&
			\multicolumn{1}{c}{$+[17dfg]$}& 
			\multicolumn{1}{c}{QED}&
			\multicolumn{1}{c}{Final}&
			\multicolumn{1}{c}{Diff. \cite{AK}}\\
			\midrule
			$2p^6   $&$^1S_0$    &        0 &   0   &   0&    0 & 0   & 0   & 0       &        \\
			$2p^5 3p$&$^3S_1$    &  6087185 &   6   & 254& 3876 & 772 & 67  & 6092159 & 0.02\% \\
			$2p^5 3p$&$^3D_2$    &  6116210 &  -21  &  24& 2886 & 701 & 43  & 6119842 & 0.03\% \\
			$2p^5 3p$&$^3D_3$    &  6129041 &  -23  &  25& 3015 & 711 & 94  & 6132864 & 0.03\% \\
			$2p^5 3p$&$^1P_1$    &  6138383 &  -11  &  41& 2825 & 704 & 82  & 6142025 & 0.03\% \\ [0.5pc]
			$2p^5 3s $ &$  2  $  &  5842248 & -10   & 108& 3408 & 735 & 787 & 5847276 & 0.03\%  \\
			$2p^5 3s $ &$  1  $  &  5857770 & -10   & 70 & 3303 & 708 & 784 & 5862626 & 0.03\%  \\
			$2p^5 3s $ &$  1  $  &  5953697 & -10   & 74 & 3364 & 717 &1042 & 5958883 & 0.03\%  \\
			$2p^5 3d $ &$^3P_1$  &  6466575 & -11   & 16 & 2384 & 665 & 87  & 6469717 & 0.03\%  \\
			$2p^5 3d $ &$^3P_2$  &  6481385 & -13   & 16 & 2250 & 658 & 86  & 6484383 & 0.03\%  \\
			$2p^5 3d $ &$^3F_4$  &  6482549 & -12   & 27 & 1745 & 622 & 97  & 6485028 & 0.03\%  \\
			$2p^5 3d $ &$^3F_3$  &  6488573 & -14   & 26 & 1740 & 607 & 84  & 6491016 & 0.03\%  \\
			$2p^5 3d $ &$^1D_2$  &  6502481 & -17   & 21 & 1696 & 627 & 88  & 6504895 & 0.03\%  \\
			$2p^5 3d $ &$^3D_3$  &  6511163 & -18   & 18 & 1762 & 604 & 87  & 6513617 & 0.02\%  \\
			$2p^5 3d $ &$^3D_1$  &  6548550 & -16   & -3 & 1747 & 616 & 134 & 6551029 & 0.02\%  \\
			$2p^5 3d $ &$^3F_2$  &  6589977 & -16   & 22 & 1729 & 629 & 335 & 6592676 & 0.02\%  \\
			$2p^5 3d $ &$^3D_2$  &  6596316 & -17   & 14 & 1947 & 641 & 334 & 6599235 & 0.03\%  \\
			$2p^5 3d $ &$^1F_3$  &  6600744 & -17   & 19 & 1803 & 610 & 343 & 6603501 & 0.03\%  \\
			$2p^5 3d $ &$^1P_1$  &  6656872 & -8    & -52& 1743 & 619 & 288 & 6659462 & 0.02\%  \\
			$3C-3D$&    (eV)	 & 13.4302  & 0.0009&-0.0061&-0.0005&0.0004&	0.0191&	13.4440& 0.15\% \\
			\bottomrule
		\end{tabular}
\end{specialtable*}

We started with all possible single and double excitations from the $2s^2 2p^6$, $2s^2 2p^5 3p$ even and $2s^2 2p^5 3s$, $2s^2 2p^5 3d$, $2s 2p^6 3p$, $2s^2 2p^5 4d$, $2s^2 2p^5 5d$ odd configurations, correlating 8 electrons with a short $[5spdf6g]$ basis set. Separate calculations were done to establish the effects of triple and quadruple excitations, as well as full correlation with the $1s^2$ shell. We found negligible corrections to both energies and matrix elements as illustrated by Table~\ref{fetable}. The differences between our theoretical energies and experimental values were found to be less than those of the NIST database by 3000 cm$^{-1}$. The energies from the revised analysis of the Fe$^{16+}$ spectra given in Table~\ref{fetable} are estimated to be accurate to about 90 cm$^{-1}$. The last line of Table~\ref{fetable} shows the difference of the $3C$ and $3D$ energies in eV, with the final value 13.44(5) eV. 

%%%%%%%%%%%%%%%%%%%%%%%%%%%%%%%%%%%%%%%%%%
\section{Conclusion and further developments}
We have developed a new parallel atomic structure code package that has opened a lot of new possibilities of high-precision calculations of atomic properties of various complex many-electron systems. The efficiency and accuracy of our programs have been validated by solving many problems, involving astrophysics, in metrology of highly charged ions, and negative ions. 

Although our parallel programs have displayed great success in efficiency and accuracy, there is still plenty of developmental work to be done. This includes optimization of current parallel algorithms, parallelization of remaining problematic serial codes (e.g. the Davidson procedure in the CI program), optimization of memory usage, completion of documentation, including check-pointing, making the codes user-friendly, and more. 

The programs developed here are part of a larger project developing an online portal for high-precision atomic data and computation. The portal will feature a database of all high-precision data calculated in the last couple of decades, as well as wave functions of large number of atoms and ions stored on a back-end server. Users will be able to request transition properties and polarizabilities for systems not in the database without having to download any codes or learn any inputs. Requested data will be generated automatically through code executions on a back-end server and then updated on the database. All codes will also be released to the public, optimized and user-friendly.  

\vspace{6pt} 
%%%%%%%%%%%%%%%%%%%%%%%%%%%%%%%%%%%%%%%%%%
%% optional
%\supplementary{The following are available online at \linksupplementary{s1}, Figure S1: title, Table S1: title, Video S1: title.}

% Only for the journal Methods and Protocols:
% If you wish to submit a video article, please do so with any other supplementary material.
% \supplementary{The following are available at \linksupplementary{s1}, Figure S1: title, Table S1: title, Video S1: title. A supporting video article is available at doi: link.}

%%%%%%%%%%%%%%%%%%%%%%%%%%%%%%%%%%%%%%%%%%
%\authorcontributions{For research articles with several authors, a short paragraph specifying their individual contributions must be provided. The following statements should be used ``Conceptualization, X.X. and Y.Y.; methodology, X.X.; software, X.X.; validation, X.X., Y.Y. and Z.Z.; formal analysis, X.X.; investigation, X.X.; resources, X.X.; data curation, X.X.; writing---original draft preparation, X.X.; writing---review and editing, X.X.; visualization, X.X.; supervision, X.X.; project administration, X.X.; funding acquisition, Y.Y. All authors have read and agreed to the published version of the manuscript.'', please turn to the  \href{http://img.mdpi.org/data/contributor-role-instruction.pdf}{CRediT taxonomy} for the term explanation. Authorship must be limited to those who have contributed substantially to the work~reported.}

\funding{This work was supported in part by U.S. NSF Grant
No. PHY-1620687 and Office of Naval Research Grant
No. N00014-17-1-2252. S. G. P. acknowledges support by the Russian Science
Foundation under Grant No. 19-12-00157}

\acknowledgments{This research was supported in part through the use of the Caviness and DARWIN HPC systems at the University of Delaware. The authors wish to thank Dr. Jeffrey Frey for helpful discussions and contributions on optimizing parts of the CI program.}

\conflictsofinterest{The authors declare no conflict of interest.}

%% Optional
%\sampleavailability{Samples of the compounds ... are available from the authors.}

%%%%%%%%%%%%%%%%%%%%%%%%%%%%%%%%%%%%%%%%%%
%% Only for journal Encyclopedia
%\entrylink{The Link to this entry published on the encyclopedia platform.}

%%%%%%%%%%%%%%%%%%%%%%%%%%%%%%%%%%%%%%%%%%
%% Optional
%\abbreviations{The following abbreviations are used in this manuscript:\\
%
%\noindent
%\begin{tabular}{@{}ll}
%UD & University of Delaware\\
%CI & Configuration interaction\\
%MBPT & Many-body perturbation theory\\
%MPI & message passing interface
%\end{tabular}}

%%%%%%%%%%%%%%%%%%%%%%%%%%%%%%%%%%%%%%%%%%
%% Optional
\appendixtitles{yes} % Leave argument "no" if all appendix headings stay EMPTY (then no dot is printed after "Appendix A"). If the appendix sections contain a heading then change the argument to "yes".
\appendixstart
\appendix

%\subsection{}
%The appendix is an optional section that can contain details and data supplemental to the main text---for example, explanations of experimental details that would disrupt the flow of the main text but nonetheless remain crucial to understanding and reproducing the research shown; figures of replicates for experiments of which representative data are shown in the main text can be added here if brief, or as Supplementary Data. Mathematical proofs of results not central to the paper can be added as an appendix.
%
%\begin{specialtable}[H]
%%\tablesize{\scriptsize}
%\caption{This is a table caption. Tables should be placed in the main text near to the first time they are~cited.\label{tab:ir17_energies}}
%%\tablesize{} % You can specify the fontsize here, e.g., \tablesize{\footnotesize}. If commented out \small will be used.
%\begin{tabular}{ccc}
%\toprule
%\textbf{Title 1}	& \textbf{Title 2}	& \textbf{Title 3}\\
%\midrule
%Entry 1		& Data			& Data\\
%Entry 2		& Data			& Data\\
%\bottomrule
%\end{tabular}
%\end{specialtable}
%
%\section{}
%All appendix sections must be cited in the main text. In the appendices, Figures, Tables, etc. should be labeled, starting with ``A''---e.g., Figure A1, Figure A2, etc.

%%%%%%%%%%%%%%%%%%%%%%%%%%%%%%%%%%%%%%%%%%
\newpage
\end{paracol}

\reftitle{References}

% Please provide either the correct journal abbreviation (e.g. according to the “List of Title Word Abbreviations” http://www.issn.org/services/online-services/access-to-the-ltwa/) or the full name of the journal.
% Citations and References in Supplementary files are permitted provided that they also appear in the reference list here.

%=====================================
% References, variant A: external bibliography
%=====================================
\externalbibliography{yes}
\bibliography{symmetry}

\end{document}